# High-pressure powder x-ray diffraction study of $EuVO_4$

Alka B. Garg[1] and D. Errandonea[2,*]

[1]High Pressure and Synchrotron Radiation Physics Division, Bhabha Atomic Research Centre, Mumbai 400085, Maharashtra, India

[2]Departamento de Física Aplicada-ICMUV, MALTA Consolider Team, Universidad de Valencia, Edificio de Investigación, C/Dr. Moliner 50, Burjassot, 46100 Valencia, Spain

**Abstract:** The high-pressure structural behavior of europium orthovanadate has been studied using *in-situ*, synchrotron based, high-pressure x-ray powder diffraction technique. Angle-dispersive x-ray diffraction measurements were carried out at room temperature up to 34.7 GPa using a diamond-anvil cell, extending the pressure range reported in previous experiments. We confirmed the occurrence of zircon-scheelite phase transition at 6.8 GPa and the coexistence of low- and high-pressure phases up to 10.1 GPa. In addition, clear evidence of a scheelite-fregusonite transition is found at 23.4 GPa. The fergusonite structure remains stable up to 34.7 GPa, the highest pressure reached in the present measurements. A partial decomposition of $EuVO_4$ was also observed from 8.1 to 12.8 GPa, however, this fact did not preclude the identification of the different crystal structures of $EuVO_4$. The crystal structures of the different phases have been Rietveld refined and their equations of state (EOS) have been determined. The results are compared with the previous experimental data and theoretical calculations.

[*]Corresponding author: daniel.errandonea@uv.es

## Introduction

Europium vanadate, $EuVO_4$, is a rare earth orthovanadate that crystallizes in the tetragonal zircon structure (space group: $I4_1/amd$, Z = 4) at ambient pressure and temperature. Zircon-type vanadates are technologically important materials with many applications [1 - 4]. Recently, these materials have attracted considerable attention due to potential applications in renewable energy and alternative green technology [5]. In addition, given their luminescent properties, chemical stability, and non-toxicity, orthovanadate nanoparticles can be used in biomedical applications [6]. Under high pressure (HP) $EuVO_4$ undergoes a structural phase transition near 8 GPa to the tetragonal scheelite-type structure (space group: $I4_1/a$, Z = 4) [7, 8] and a subsequent transition at 21 GPa to a phase that is proposed to have a monoclinic fergusonite-type structure (space group: $I2/a$, Z = 4) [7]. During the last years, several studies have been carried out on the HP behavior of orthovanadates, showing that pressure is an efficient tool to improve the understanding of their physical properties [7, 27]. In particular, x-ray diffraction (XRD) [7, 9-15], optical [8, 16, 17], and Raman scattering measurements [18-22] as well as theoretical studies [10, 14 , 22-27] have been carried out to understand the effects of pressure on orthovanadates. From these studies, it was determined that vanadates with small rare-earth cations (Sm and those atoms with smaller ionic radii) transform at HP to the tetragonal scheelite-type structure while those with large cations (Pr and those atoms with a larger ionic radii) transform to the monoclinic monazite-type structure (space group: $P2_1/n$, Z = 4).

In spite of the studies summarized above, there are several issues that remain unresolved on the HP behavior of compounds like $EuVO_4$. One of them is the compressibility of the zircon structure, for which values ranging from 118 GPa to 172 GPa have been reported for the bulk modulus [7, 16, 24 - 26]. A second issue that deserves to be addressed is the influence of deviatoric stresses on the HP structural sequence; these stresses have been shown to considerably influence the HP structural behavior of the isomorphic $HoVO_4$ and $CeVO_4$ [14, 28]. Other facts that should be explored in detail are the possible decomposition or amorphization of $EuVO_4$. Partial decomposition was previously detected under compression in $HoVO_4$ [14] and pressure-driven amorphization has been reported to occur in related scheelite-type oxides [29]. Finally, in spite of the fact that the scheelite-type $EuVO_4$ has been obtained from different experiments [7, 30], its crystal structure has not been yet fully refined. The same can be stated for the crystal structure of the second HP phase. The crystal structure of the scheelite-type structures of vanadates has been calculated for isostructural compounds [10, 31] but not for $EuVO_4$.

In order to shed more light on the understanding of the HP structural properties of $EuVO_4$, we have studied this compound by powder XRD up to 34.7GPa using Elettra synchrotron radiation source. Under quasi-hydrostatic conditions we confirmed the zircon-to-scheelite transition, whose onset is found at 6.8 GPa. The structural details of the zircon and scheelite phases have been obtained by Rietveld refinements. We also observed that in the presence of deviatoric stresses, a second transition takes place at 23.4 GPa. This HP phase also has been Rietveld refined and assigned to the previously proposed monoclinic fergusonite-type structure [7]. In the present experiments, a minor partial decomposition of $EuVO_4$ was also observed, which is probably triggered by the absorption of x-rays by the sample. The obtained results are compared with previously reported experimental and theoretical studies. The axial compressibilities and room temperature P-V equation of state (EOS) of the different phases of $EuVO_4$ are also reported.

**Experimental Details**

The compound, $EuVO_4$, used in the present experiments was prepared by solid-state reaction of appropriate amounts of $Eu_2O_3$ (Alfa Aesar 99.9 %) and $V_2O_5$ (Aldrich 99.6%). The constituent oxides were heated at 425 K for 24 hours prior to weighing to remove moisture or any other organic impurity. The reactants were thoroughly ground with an agate mortar and pestle for two hours. Pellets of 12.5 mm diameter and 5 mm thickness were made from this powder by cold pressing. These pellets were heated at 800 °C for 24 hrs in chamber furnace and cooled to room temperature. The as prepared sample was characterized for its single phase formation by powder x-ray diffraction (XRD) using rotating anode generator (Rigaku-Make) operating at 50 kV and 50 mA current with the molybdenum (Mo) $K_\alpha$ ($\lambda$=0.7107 Å) radiation. A highly oriented pyrolytic graphite monochromator with (002) plane orientation is used for selecting the $K_\alpha$ radiation of Mo. A single phase with the zircon-type structure was confirmed for $EuVO_4$ with unit-cell parameters $a$ = 7.2408(5) Å and $c$ = 6.3681(3) Å, which agrees well with the values reported in literature [32].

High-pressure experiments were performed at room temperature up to 34.7 GPa using 4:1 methanol-ethanol as pressure-transmitting medium (PTM) [33, 34]. Angle-dispersive XRD experiments were carried out using a diamond-anvil cell (DAC) equipped with diamond culets of 400 μm, having large angular aperture to favor accurate structural refinements. The pressure chamber was a 150 μm hole drilled in a hardened stainless-steel gasket, pre-indented to a thickness of 60 μm. The sample was loaded in the pressure chamber together with a few platinum (Pt) grains. Pressure was determined with an accuracy of 0.05 GPa using the EOS of Pt [35]. *In-situ* HP-XRD measurements were carried out at the XRD1 beam-line of Elettra synchrotron source. Monochromatic x-rays ($\lambda$ = 0.5997 Å) were used with the beam-size limited to 80 μm in diameter using a circular collimator. Images of the powder diffraction rings were collected on a MAR345 image-plate area detector. Exposure times of 15–20 minutes were employed at each measurement. The two dimensional diffraction images collected were integrated using FIT2D [36].

Structural analysis was performed with GSAS [37]. For a typical refinement, first the background was fitted with a Chebyshev polynomial function of first kind with six or eight coefficients depending on the shape of the background in the diffraction pattern. The Bragg peak profiles were modelled using a pseudo-Voigt function with total eighteen coefficients. Since the occupancy and the atomic displacement factors are correlated and more sensitive to background determination than the positional parameters, they were constrained to 1 and B = 0.025 $Å^2$, where B is the overall displacement factor, to reduce the number of free parameters used in the refinement. The next step was to refine the unit cell parameters followed by the refinement of atomic positional parameters. Finally a full refinement was carried out. In those cases were a multiphase refinement was required, independent peaks were carefully selected to determine the initial values of the relative scale factors.

**Results and Discussion**

Fig.1 shows XRD patterns measured up to 10.1GPa. Five Bragg peaks belonging to Pt (used as *in-situ* pressure calibrant) are labeled on it. They can be easily identified since diffraction peaks from Pt have different pressure evolution than those of the sample. The diffraction peaks from the sample in XRD patterns collected from ambient pressure to 6 GPa, could be unequivocally assigned to the zircon structure. This is illustrated in Fig. 2(a) by the Rietveld refinement of the diffraction pattern measured at 1.8 GPa, for which the residuals of the structural refinement are also shown. The R-factors obtained from the refinement are $R_p$ = 4.68%, $R_{wp}$ = 6.21% and the reduced $\chi^2$ =

2.037. These reliability factors are comparable to those obtained in previous studies in isostructural compounds [11, 14]. The unit-cell parameters determined at 1.8 GPa are *a* =7.2133(3) Å and *c*=6.3559(5) Å. Table I gives the refined atomic positions. Similar qualities of refinements were obtained up to 6 GPa. Unit-cell parameters have been determined at all pressures and the effect of pressure on them will be discussed below. Atomic positions have also been determined at all pressures. The only two positions that are not fixed by the symmetry gradually change with pressure shifting the *y* coordinate of oxygen atoms from 0.4303(18) at 1.8 GPa to 0.4367(23) at 6 GPa and the *z* coordinate of oxygen atoms from 0.2095(19) at 1.8 GPa to 0.2057(24) 6 GPa; i.e. the change induced by pressure in the atomic positions up to 6 GPa is comparable with their uncertainties.

When pressure reaches 6.8 GPa, a new Bragg peak emerges near 2θ = 12º. This peak is identified by the symbol $ in Fig. 1. Upon compression, this and other extra peaks gradually grow in intensity and simultaneously the peaks indexed as zircon phase of $EuVO_4$ gradually vanish. The highest pressure at which the most intense peak of the zircon phase is observed is 10.1 GPa. These changes can be assigned to the onset of a phase transition, with the zircon and first HP phase coexisting between 6.8 and 10.1 GPa. In Fig. 2(b) we show the refinement of the XRD pattern measured at 8.1 GPa to illustrate the coexistence of zircon and scheelite structures. The various R-factors of the refinement are $R_p$ =3.60%, $R_{wp}$ = 5.09% and the reduced $\chi^2$ = 12.5. The HP phase finally appears as a single phase at 11.4 GPa. The onset pressure of the transition, 6.8 GPa, is close to that previously found in XRD measurements carried out using silicone oil as PTM [7].The pressure where pure scheelite is observed in the present measurements is around 4 GPa lower than the reported in earlier studies [7].

Fig. 3 shows the XRD patterns measured upon compression from 11.4 to 20.8 GPa. The Rietveld refinements of all these patterns confirmed that the patters measured in the above given pressure range can be assigned to a scheelite-type structure. Thus, it can be undoubtedly stated that the crystal structure of the HP phase is isomorphic to scheelite. This crystal structure was previously identified by means of a LeBail (profile refinement) method [7]. Therefore, the present study reports the complete structure of the scheelite-type HP phase of $EuVO_4$ for the first time. The residuals of the Rietveld refinement carried out for the data collected at 15.3 GPa are shown in Fig. 2(c). The R-factors obtained from the refinement are $R_p$ = 3.52%, $R_{wp}$ = 5.02% and the reduced $\chi^2$= 22.5. The unit-cell parameters of scheelite-type $EuVO_4$ at 15.3GPa are $a$ = 5.004(2) Å and $c$ = 11.128(5) Å. Table I gives the refined atomic positions for scheelite-type $EuVO_4$ at 15. 3 GPa. The structure of scheelite-type $EuVO_4$ is similar to that of other scheelie-type vanadates [9, 10, 11, 14, 27] determined either from experiments or calculations. It is also in good agreement with the unit-cell parameters determined from a previous XRD study of $EuVO_4$ [7].

It is important to note that in addition to the peaks assigned to zircon, and scheelite $EuVO_4$ and Pt, there are a few weak peaks detected in the data set collected from 8.1 to 12.8 GPa (see Figs. 1 and 3). The most intense of them is located near 2θ = 11° and is denoted by an asterisk in Figs. 1 and 3. These peaks are not consistent with a possible monazite structure of $EuVO_4$, which has been found as a post-zircon phase in lighter rare-earth vanadates; *e.g.* $PrVO_4$ and $CeVO_4$ [9, 13]. In contrast, we found that the observed extra peaks can be accounted by an orthorhombic structure with the space group, *Pmmn*, and unit-cell parameters of $V_2O_5$ [38]. However due to low intensity of the peaks assigned to $V_2O_5$, a full structural refinement of its structure could not be carried out.  The assignment of the extra peaks to $V_2O_5$ can be done at all the pressures where these peaks have been observed. Therefore, the possibility of pressure-induced partial decomposition of $EuVO_4$ in the experiments is a probable hypothesis to explain our observations. This hypothesis is consistent with the partial decomposition of $HoVO_4$ found under similar experimental conditions [14]. We have also observed similar effects in ongoing studies of other orthovanadates like $SmVO_4$ and $DyVO_4$. Fortunately, the partial decomposition of the sample and the Bragg peaks associated with it (which are weak suggesting the decomposition of a minor fraction of the sample) do not preclude the identification of the zircon and scheelite structures at any pressure. The Bragg peaks of either one of these two structures are always the dominant peaks in all the XRD patterns. Additionally, at 13.7 GPa the peaks assigned to $V_2O_5$ disappear from the XRD pattern. This observation is explained by the fact that pressure induces pronounce structural disorder in $V_2O_5$ [39], which leads to broadening and weakening of Bragg peaks in $V_2O_5$ and ultimately to the amorphization of $V_2O_5$. This amorphization is irreversible hence no Bragg peaks associated to $V_2O_5$ are expected upon pressure release. The cause of the partial decomposition of $EuVO_4$ is beyond the scope of this study. One possibility is that the partial decomposition could be triggered by the x-ray absorption which could induce photoelectric processes leading to the dissociation of $V_2O_5$ units as proposed in a previous study on $HoVO_4$ [14]. A similar decomposition has been recently observed in ternary oxides also under HP conditions [40], in particular when x-rays with wavelengths similar to those used in the present experiments are employed.

As shown in Fig. 4, on further compression, we found that starting at 23.4 GPa the Bragg peaks of scheelite $EuVO_4$ considerably broaden. Additionally, extra weak peaks appear. In particular, the strongest peak of scheelite assigned to the (112) and (013) Bragg peaks of scheelite, which is near $2\theta = 11^{\circ}$, develops a shoulder on the right-hand side becoming asymmetric. Also the (011) Bragg reflection of scheelite (located near 2θ = 7.5º) splits into the (110) and (011) Bragg peaks of the new structure, and there is an additional peak appearing at low angles, which is highlighted by a dotted line in Fig. 4 that follows it as pressure changes. Similar changes have been observed in HP x-ray diffraction studies on other rare-earth orthovanadates [7, 14, 41]. These changes are consistent with a transformation from the scheelite-type structure to a monoclinic fergusonite-type structure as previously observed at a similar pressure in $EuVO_4$ [7]. Since in the previous study the fergusonite structure was proposed based on a LeBail analysis (no Rietveld refinement) [7], the structural details of the fergusonite phase, extracted from the Rietveld refinement of the present data, are being reported for the first time. We found that the fergusonite structure of $EuVO_4$ remains stable up to 34.7GPa, the highest pressure reached in the present experiments. The unit-cell parameters of fergusonite-type $EuVO_4$ at 25.6 GPa are *a* = 5.036 (4) Å, *b* = 11.077(10) Å, *c* = 4.675(6) Å, and β = 92.79(6)º. The residuals of the refinement carried out for the data collected at 25.6 GPa are shown in Fig. 2(d). The R-factors of the refinement are $R_p$= 2.20%, $R_{wp}$ = 3.39% and the reduced $\chi^2$= 4.26. Comparable quality of refinement was obtained up to 32.9 GPa. For the data collected at 34.7 GPa, the quality of the refinement was worsened (may be due to the lower counts we had for this pressure) and therefore structural parameters have not been obtained at this pressure. However, the XRD pattern measured at 34.7 GPa could still be indexed with the fergusonite structure. Table I gives the atomic positions obtained for the fergusonite structure at 25.6 GPa. The reported atomic positions are consistent with those calculated for fergusonite-type $TbVO_4$ [18]. Note that the structural refinements for the fergusonite structure have been done without imposing any constrain to the oxygen coordinates, but taking as starting values those derived from scheelite by means of the group-subgroup relationship between scheelete and fergusonite ($I2/a$ is a maximal subgroup of $I4_1/a$) [42]. Our measurements confirm the existence of fergusonite-type HP polymorph of $EuVO_4$. In addition, crystal-structure data of the HP fergusonite phase of $EuVO_4$ have been obtained as a function of pressure for the first time. Upon decompression the scheelite-fergusonite transition is reversible. When decreasing pressure from 34.7 GPa we observed the fergusonite structure at 29.8, 23.9 and 14.4 GPa. In a subsequent decompression step, the scheelite phase was recovered at 7.6 GPa. In contrast, the zircon-scheelite transition has been previously determined to be a nonreversible transformation [7], a fact that we confirmed.

Before discussing the pressure effects on unit-cell parameters and volume we would like to mention a few other observations. First one is that no additional decomposition was detected in the diffraction patterns collected in our experiments for pure scheelite-type and fergusonite-type $EuVO_4$. This indicates that pressure-induced decomposition is not prevalent in $EuVO_4$ up to 34.7 GPa. The other observations are that we did not find either the occurrence of pressure-induced amorphization or color changes in the sample that may indicate pressure-induced metallization as found in $CeVO_4$ at 11 GPa [12]. We would also like to comment here that it was previously observed, in related compounds including $TbVO_4$ [13, 18], that the scheelite-fergusonite transition pressure depends upon the PTM used in the experiments. It is well known that the use of different PTM could induce different deviatoric stresses within the pressure chamber of the DAC leading to conditions that range from quasi-hydrostatic to non-hydrostatic. This could strongly influence the HP structural sequence of the materials even at low pressures [34, 43, 44]. This influence is prevalent in scheelite-type oxides [45, 46]. In our case, the transition pressure of scheelite-fergusonite phase transition is similar with that obtained in the experiments performed using silicon oil as PTM [7], which suggest that deviatoric stresses are similar in both the experiments. The Bragg peaks were observed to broaden gradually after the completion of the zircon-scheelite transition in the present data as also reported in Ref. 7. This fact indicates that a certain degree of non-hydrostaticity is present in the experiments and therefore the influence of deviatoric stresses in the results cannot be neglected.

The pressure dependence of unit-cell parameters of the three different phases of $EuVO_4$ are extracted from the Rietveld refined data and are plotted in Fig. 5. In Fig. 6 we show the pressure dependence of the unit-cell volume for zircon, scheelite, and fergusonite phases of $EuVO_4$. The obtained pressure dependence of the unit-cell parameters for the zircon phase is similar to that obtained in Ref. 7. Experiments for the pressure range of stability of the zircon phase can be considered quasi-hydrostatic, because the Bragg peaks do not become broader up to the highest pressure where the zircon phase is present. The pressure evolution of the unit-cell parameters of zircon-type $EuVO_4$ can be described by a linear function, with axial compressibilities $\kappa_a = 2.5 \times 10^{-3}$ $GPa^{-1}$ and $\kappa_c = 1.3 \times 10^{-3}$ $GPa^{-1}$. The fact that the *a*-axis of zircon $EuVO_4$ is more compressible that the *c*-axis, is a typical behavior of zircon-structured oxides [47], in particular vanadates [7, 14]. The origin of this behavior is related with the packing of $EuO_8$ and $VO_4$ polyhedral units in the zircon structure. This structure can be considered as chains of alternating edge-sharing $VO_4$ tetrahedra and $EuO_8$ dodecahedra extending parallel to the *c*-axis with the chains joined along the *a*-axis by edge-sharing $EuO_8$ dodecahedra. As we will show later, in $EuVO_4$ the $VO_4$ units are much less compressible than the $EuO_8$ units. This makes the *c*-axis less compressible than the *a*-axis as observed in our experiments.

For the scheelite-structure, the pressure behavior found for the unit-cell parameters is also similar to the one reported in Ref. 1 up to 15.3 GPa. Beyond this pressure, we observed that the compressibility of the sample is highly reduced, which deviates with the results reported in Ref. 7. This fact is more noticeable for the *c*-axis than for the *a*-axis. This occurs together with a gradual broadening of XRD peaks. Since the PTM used in the present experiment is expected to induce better quasi-hydrostatic conditions than the silicone oil used in Ref. 7 [34], we believe that the results we observed beyond 15.3 GPa could be influenced by sample bridging with the diamond anvils [44]. Sample bridging usually induces large deviatoric stresses and could be caused by the reduction of the gasket thickness that occurs under compression. Therefore, we will limit the discussion of the compressibility of scheelite-type $EuVO_4$ to the data collected for $P \leq 15.3$ GPa. The unit-cell parameters of scheelite-type $EuVO_4$ can also be described by a linear function in the pressure range 8.1 – 15.3 GPa, being the axial compressibilities $\kappa_a = 1.3 \times 10^{-3}$ $GPa^{-1}$ and $\kappa_c = 1.9 \times 10^{-3}$ $GPa^{-1}$. In this case, the most compressible axis is the *c*-axis. Since the scheelite and zircon structure are made of similar polyhedral units [47], the anisotropic behavior of scheelite can be explained using a similar argument which is used to explain the anisotropic compressibility of the zircon structure [45].

For the sake of completeness, in Fig. 5, we have included the results for fergusonite phase as well. The increase of the $\beta$ angle under compression (see inset of Fig. 5) and the increase of the splitting between the *a*-axis and *c*-axis indicate that the monoclinic distortion of fergusonite gradually increases under pressure. This phenomenon should be reflected in an enhancement of the spontaneous strains that characterize the distortion caused by the transformation from scheelite to fergusonite [48]. As a consequence of it, the splitting of Bragg reflections of fergusonite (that in the tetragonal scheelite structure gives the same peak) is gradually increased.

We have analyzed the P-V data obtained for zircon and scheelite $EuVO_4$, shown in Fig. 6, using a third-order Birch-Murnaghan (BM) EOS [49] and the EosFit7 software [50]. The EOS fit has not been performed for fergusonite $EuVO_4$ because the results obtained for this phase are probably influenced by large deviatoric stresses as commented in previous paragraph (due to sample bridging with the diamond anvils and the used PTM). From the EOS fit, we obtained the ambient pressure bulk modulus $B_0$ and its pressure derivative $B_0'$ as well as the unit-cell volume $V_0$. The EOS parameters are given in Table II together with the implied values for the second pressure derivative of the bulk modulus, $B_0'' = -\left[(B_0' - 4)(B_0' - 3) + \frac{35}{9}\right]/B_0$ [49, 51]. To fit the experimental P-V results of scheelite $EuVO_4$, we have only used the data determined for P ≤ 15.3GPa to minimize the influence of deviatoric stresses in the fits. Note that for P > 15.3 GPa, the compressibility of scheelite-type $EuVO_4$suddenly decreases. This fact is a consequence of the increase of deviatoric stresses [45]. In all the EOS fits we assumed the three EOS parameters ($V_0$, $B_0$, $B_0'$) as fitting parameters. The reported values for the EOS parameters of the different phases of $EuVO_4$ are very similar to those previously found experimentally in $EuVO_4$ [7]. They also compare well with the systematics observed in other vanadates [7, 10, 11, 13, 14]. In contrast, when comparing with other experiments, the bulk modulus determined here for the zircon phase is 25% larger than the experimental bulk modulus reported in Ref. 15. Further, density-functional theory (DFT) gives a larger bulk modulus than experiments, $B_0$= 172 GPa [24], whereas estimations using semi-empirical models [7] or the dielectric chemical-bond method [25] underestimate the value of $B_0$ at least by 15 GPa. However, calculations carried out using an electronegativity based model are the theoretical results that better agree with experiments, giving $B_0$ = 140 GPa [26]. The difference observed between the *ab initio* based calculated [24] bulk modulus and the values obtained from the experiments need to be discussed. *Ab initio* calculations and DAC experiments show an excellent agreement for $HoVO_4$ [14] and $DyVO_4$ [10]. A possible reason for this discrepancy found in $EuVO_4$ could be related to the inclusion of a Hubbard term (U) in the calculations [24]. This term is needed to better describe the strongly correlated europium electronic states, however, calculations must be performed very carefully since results strongly depend upon the value used for U [52].

On the other hand, when comparing the zircon and scheelite structure, it can be seen that the scheelite phase has larger bulk modulus than the zircon phase (see Table II). This fact is in agreement with the large volume collapse associated to the zircon-scheelite transition and to the related increase of packing efficiency of the scheelite structure. In this study, the volume collapse at the transition is $\Delta V/V \approx -10\%$, which agrees with the unit-cell volume reported for scheelite-type $EuVO_4$ in two previous works [7, 30]. Regarding the fergusonite structure, from Fig. 5 it appears that there is no volume discontinuity at the scheelite-fergusonite transition. It can also be seen that the fergusonite-type HP phase of $EuVO_4$ is more compressible than scheelite-type $EuVO_4$. However, we prefer to be cautious regarding this point and not to make any strong statement on the compressibility of fergusonite-type $EuVO_4$, because we measure it under non-hydrostatic conditions.

To conclude, we briefly comment on the effect of pressure in the polyhedral compression of $EuO_8$ and $VO_4$ units. This information can be extracted from our structural refinements. Polyhedral volumes have been calculated using Vesta [53]. We found that polyhedral volumes for both zircon- and scheelite-type vary smoothly with pressure. For the zircon structure a polyhedral bulk modulus of 250(10) GPa is obtained for the $VO_4$ tetrahedron and a polyhedral bulk modulus of 140(10) GPa is obtained for the $EuO_8$ dodecahedron. These values are estimated by using a second-order Birch-Murnaghan EOS [49]. Thus, given the small size of the $VO_4$ tetrahedron (volume 2.76 $Å^3$ at ambient pressure), compared with the $EuO_8$ dodecahedron (volume 24.97 $Å^3$ at ambient pressure), and the relative rigidity of the $VO_4$ tetrahedron, most of the volume compression in $EuVO_4$ is accounted by the Eu-O bonds. An analogous situation is observed in scheelite-type $EuVO_4$. The anisotropic compressibility of both the phases may be understood by the different compressibility of $EuO_8$ and $VO_4$ polyhedral units and the way they are interconnected as discussed above. A similar behavior has been found in scheelite-type $EuWO_4$ [52].

**Conclusions**

We performed room-temperature angle-dispersive XRD measurements on $EuVO_4$ up to 34.7 GPa. The onset of the irreversible zircon-scheelite transition was found at 6.8 GPa. Evidence of a second pressure-induced transition is detected at 23.4 GPa. This transition is from scheelite to fergusonite and is reversible, being the scheelite structure recovered upon decompression from 34.7 GPa at 7.6 GPa. In addition, a partial decomposition of $EuVO_4$ was observed from 8.1 to 12.8 GPa. We believe, it could be triggered by x-ray absorption when a large x-ray wavelength is used. The structure of the different phases have been Rietveld refined at different pressures, and the pressure dependence of the unit-cell parameters have been extracted as well as the equations of state of different phases determined. The reported results are compared with previous studies and the bulk and axial compressibilities are discussed in terms of polyhedral compressibilities.

**Acknowledgements**

This work was partially supported by the Spanish MINECO under Grant MAT2013-46649-C04-01 and by Generalitat Valenciana and Grant No. ACOMP/2014/243.The authors thank Elettra synchrotron for providing beam-time for the XRD experiments. Alka B. Garg acknowledges the DST of India for travel support and Italian Government for hospitality at Elettra. She also acknowledges fruitful discussions with Dr. Surinder M. Sharma.

**Table I.** Refined atomic positions from XRD patterns for zircon (top), scheelite (center), and fergusonite(bottom) phases of $EuVO_4$ at 1.8, 15.3, and 25.6GPa, respectively.

| Atom | Site | x | y | Z |
|---|---|---|---|---|
| Eu | 4a | 0 | 3/4 | 1/8 |
| V | 4b | 0 | 1/4 | 3/8 |
| O | 16h | 0 | 0.4303(18) | 0.2095(19) |
| Eu | 4b | 0 | 1/4 | 5/8 |
| V | 4a | 0 | 1/4 | 1/8 |
| O | 16f | 0.2653(31) | 0.399(5) | 0.4542(19) |
| Eu | 4e | 1/4 | 0.6327(10) | 0 |
| V | 4e | 1/4 | 0.1202(29) | 0 |
| O | 8f | 0.871(9) | 0.906(5) | 0.254(15) |
| O | 8f | 0.448(4) | 0.219(6) | 0.866(17) |

**Table II.** EOS parameters for different structures determined from present experiments[a] and from Ref.1[b].The pressure medium is indicated (ME: methanol-ethanol; S: silicone oil). $B_0''$ is the implied value of the second-pressure derivative of the bulk modulus (see text).

| Phase | PTM | $V_0$ (Å$^3$) | $B_0$ (GPa) | $B_0'$ | $B_0''$ (GPa$^{-1}$) |
|---|---|---|---|---|---|
| zircon[a] | ME | 333.84(4) | 150(7) | 5.3(6) | -0.0459 |
| zircon[b] | S | 333.2(9) | 149(6) | 5.6(6) | -0.0540 |
| scheelite[a] | ME | 296(2) | 195(8) | 5.5(9) | -0.0392 |
| scheelite[b] | S | 299.4(9) | 199(9) | 4.1(9) | -0.0201 |

**Figurecaptions**

**Figure 1.** Selection of XRD patterns measured up to 10.1 GPa. Pressures are indicated. The Bragg peaks of Pt are identified and the indices of most relevant peaks of the zircon phase are shown. The symbol $ identifies the most intense peak of scheelite in the pressure range where this phase coexists with zircon. The symbol * identifies the most intense peak associated with $V_2O_5$.

**Figure 2.** Rietveld refinements for different phases of $EuVO_4$; (a) zircon (1.8GPa), (b) zircon + scheelite (8.1 GPa), (c) scheelite (15.3GPa), and (d) fergusonite (25.6GPa). Dots correspond to experiments; the refinements (residuals and background) are shown with red (black) solid lines. Bragg peaks positions of $EuVO_4$and Pt are indicated by black and red ticks, respectively. In (b) Z and S are used to identify calculated reflections of zircon and scheelite.

**Figure3.** Selection of XRD patterns measured from 11.4 to 20.8 GPa. Pressures are indicated and Pt peaks are identified. The symbol * identifies the most intense peaks associated with $V_2O_5$.The indices of the most relevant peaks of the scheelite phase are shown.

**Figure 4.** Selection of XRD patterns measured from 23.4 to 34.7 GPa. Pressures are indicated and Pt peaks are identified. The (011) and (011) peaks of the fergusonite phase are shown (see text for discussion).

**Figure 5.** Unit-cell parameters versus pressure. Solid symbols: present work. Empty symbols: Refs. 1. Squares are used for zircon, circles for scheelite, and tringles for fergusnoite. The inset shows the evolution of the β angle in the fergusonite phase.

**Figure 6.** Unit-cell volume versus pressure. We show results from this experiment (solid symbols) and from Refs. 7 and 30 (empty symbols) as well as and the fitted EOS for zircon and scheelite. Squares, circles, and triangles are used for zircon, scheelite, and fergusonite, respectively. The empty diamond is from Ref. 30.

**Figure 1**

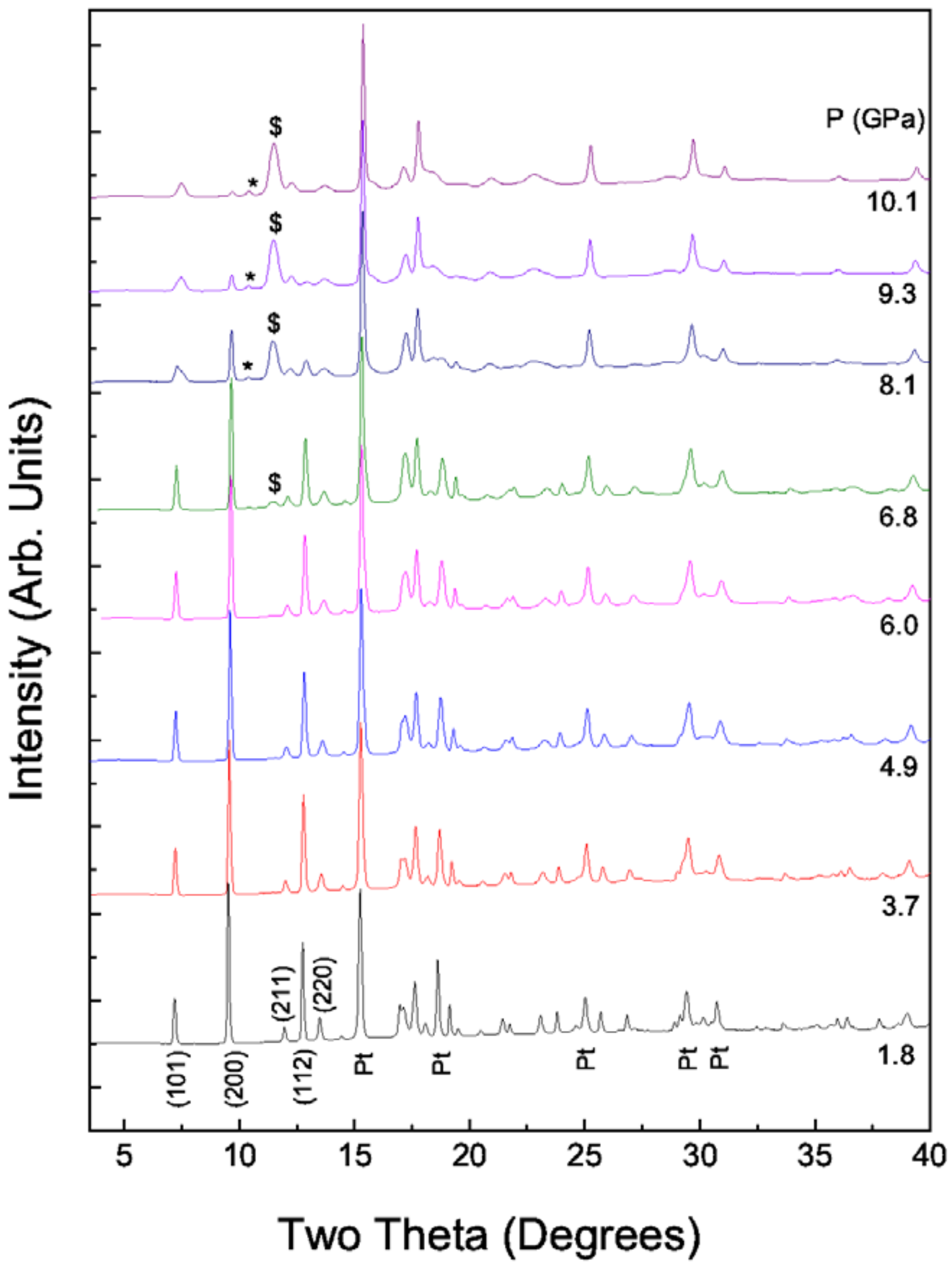

Intensity (Arb. Units)
Two Theta (Degrees)
P (GPa)
10.1
9.3
8.1
6.8
6.0
4.9
3.7
1.8
$
*
(101)
(200)
(211)
(112)
(220)
Pt
5
10
15
20
25
30
35
40

**Figure 2**

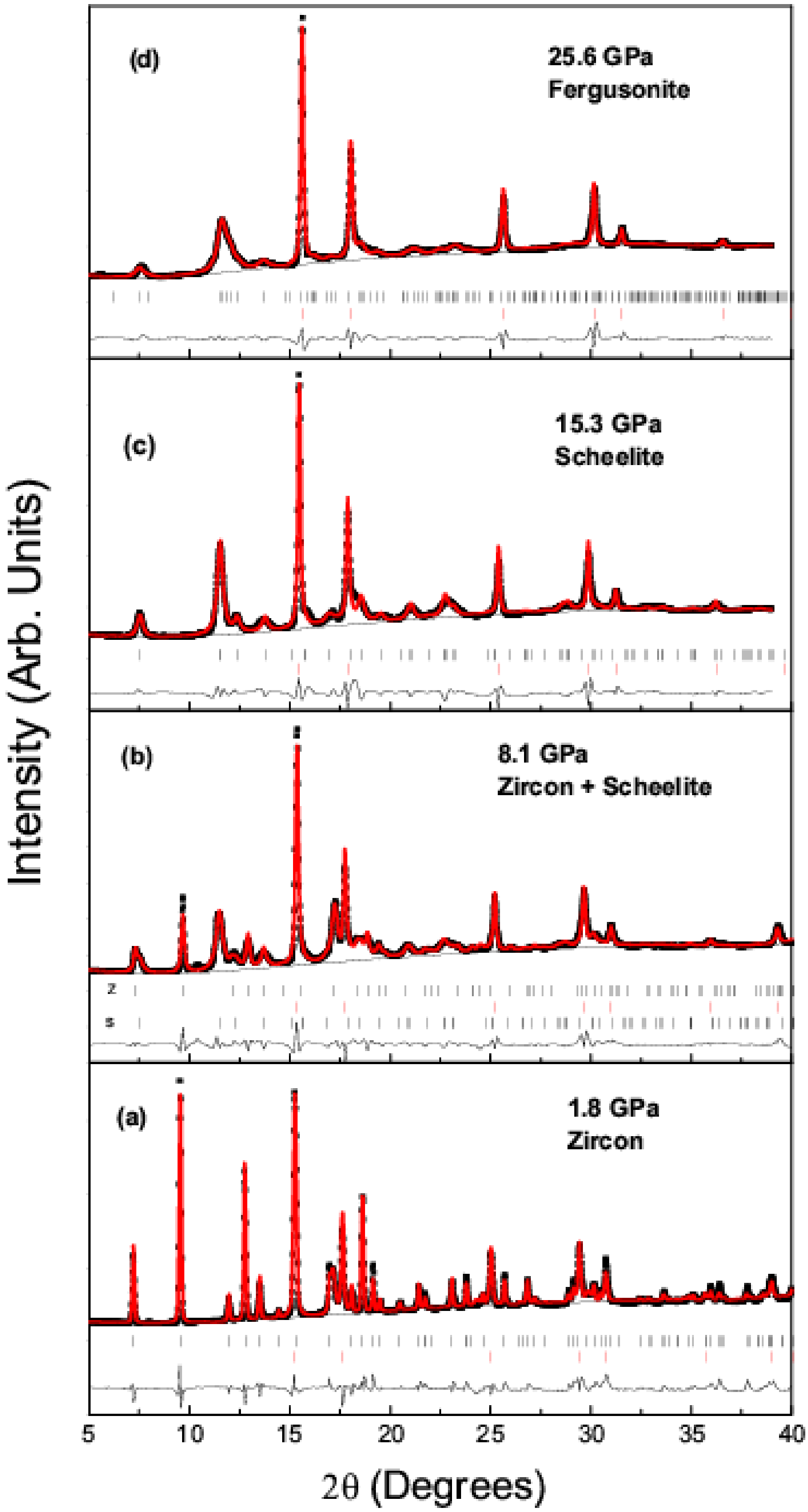
(d)
25.6 GPa
Fergusonite
(c)
15.3 GPa
Scheelite
(b)
8.1 GPa
Zircon + Scheelite
z
s
(a)
1.8 GPa
Zircon
Intensity (Arb. Units)
5
10
15
20
25
30
35
40
2θ (Degrees)

**Figure 3**

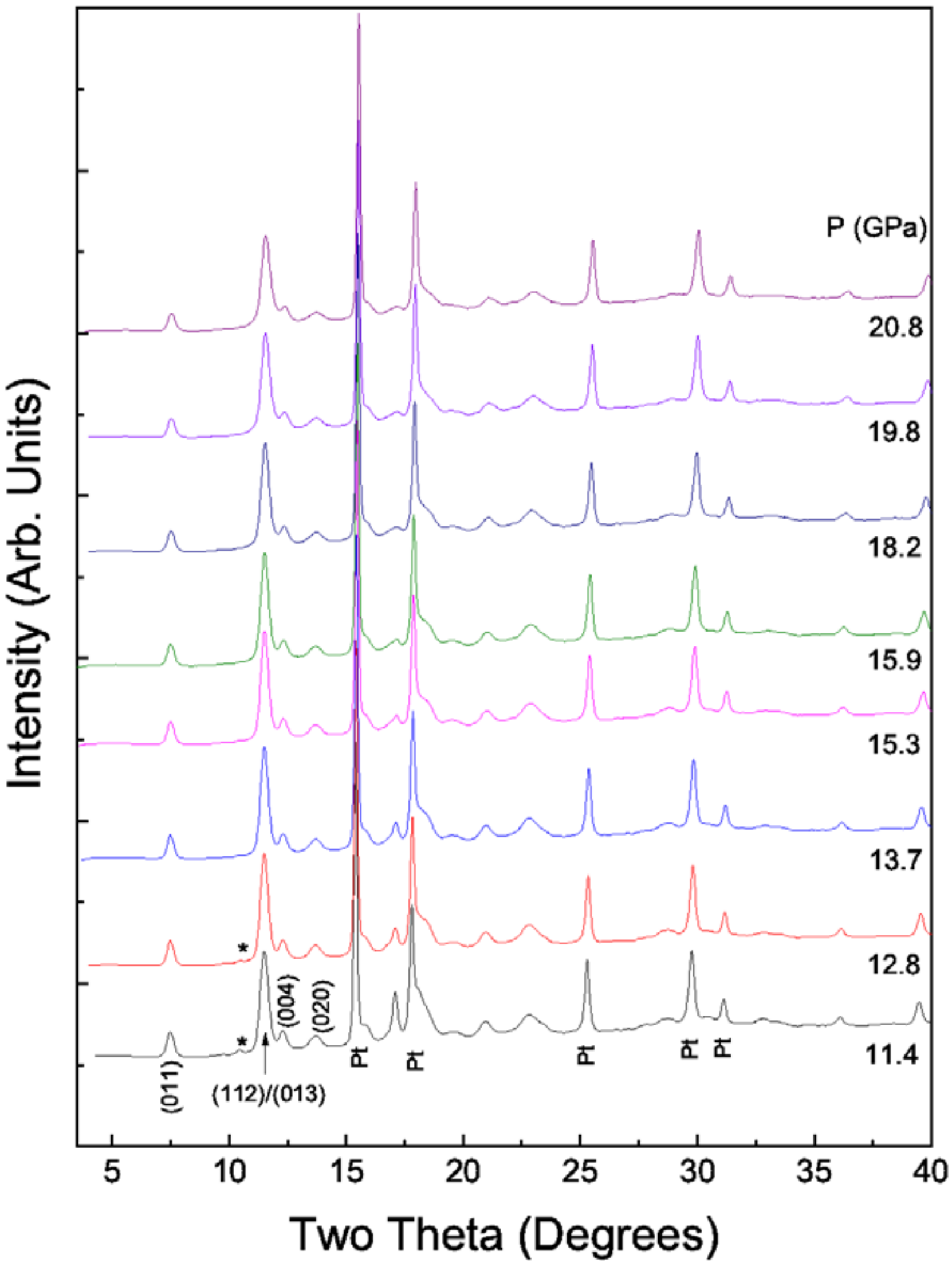

Intensity (Arb. Units)
Two Theta (Degrees)
P (GPa)
20.8
19.8
18.2
15.9
15.3
13.7
12.8
11.4
(011)
(112)/(013)
(004)
(020)
Pt
Pt
Pt
Pt
Pt
*
*
5
10
15
20
25
30
35
40

**Figure 4**

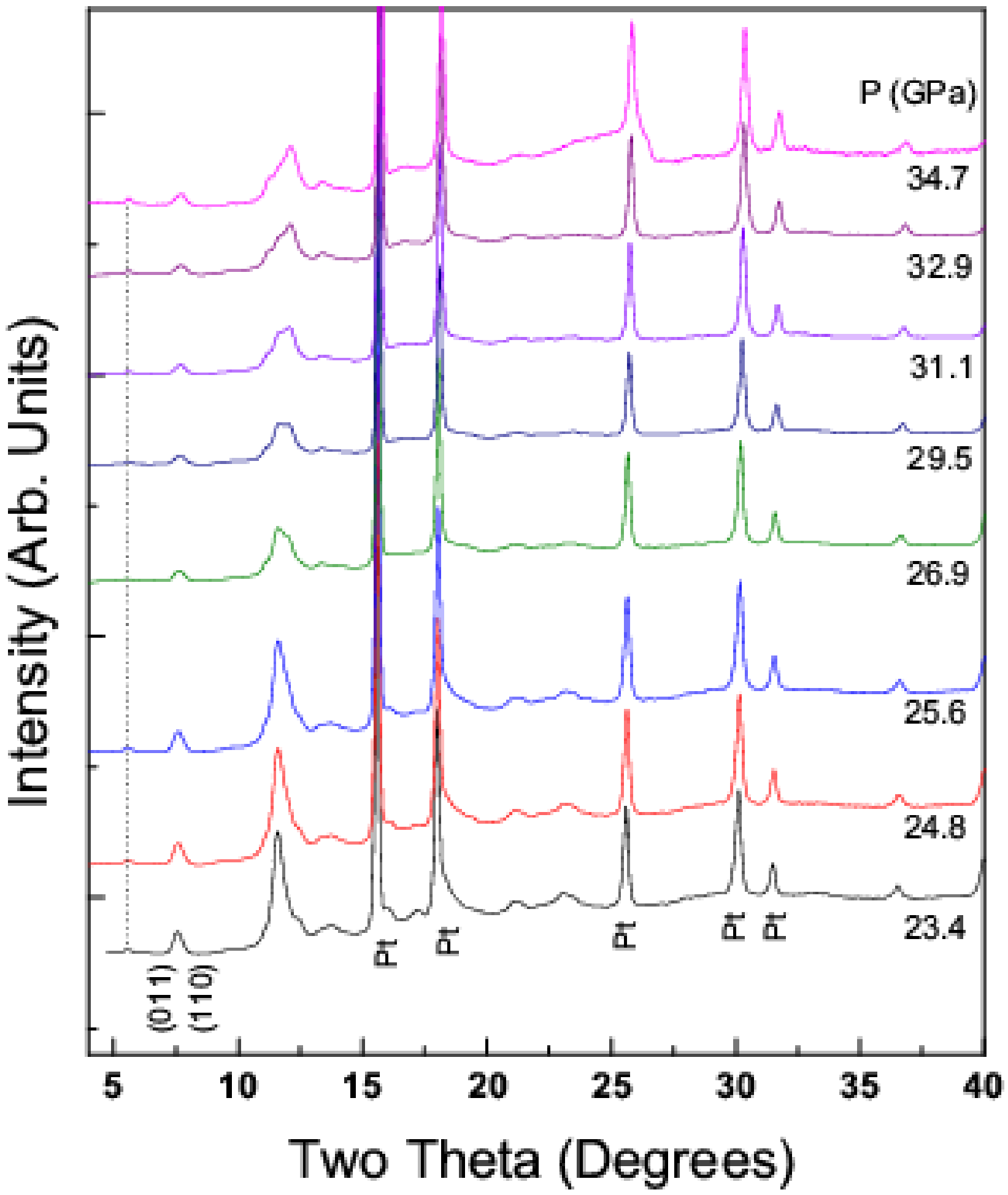

P (GPa)
34.7
32.9
31.1
29.5
26.9
25.6
24.8
23.4
Pt
Pt
Pt
Pt
Pt
(011)
(110)
Intensity (Arb. Units)
5
10
15
20
25
30
35
40
Two Theta (Degrees)

**Figure 5**

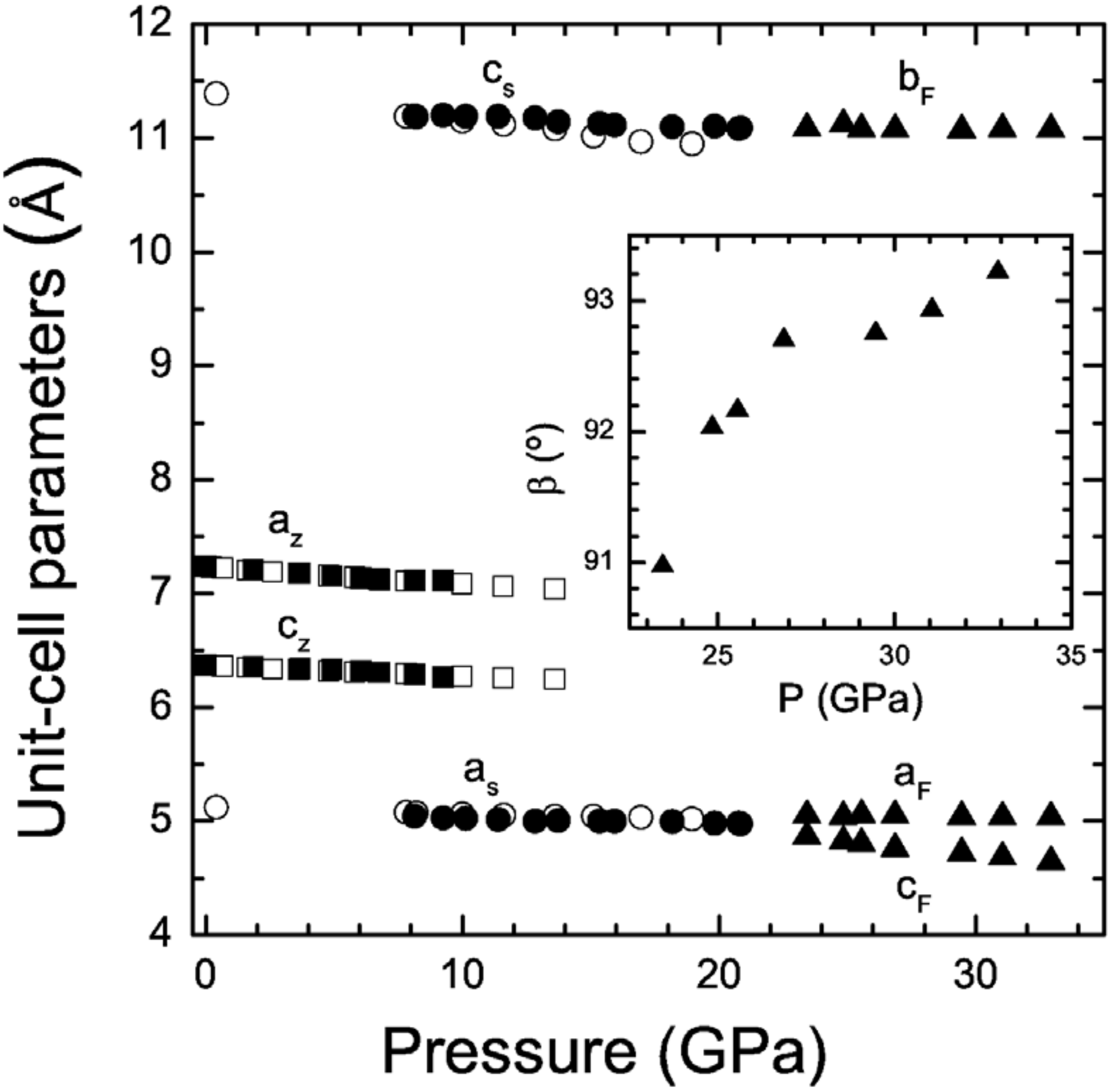

**Figure 6**

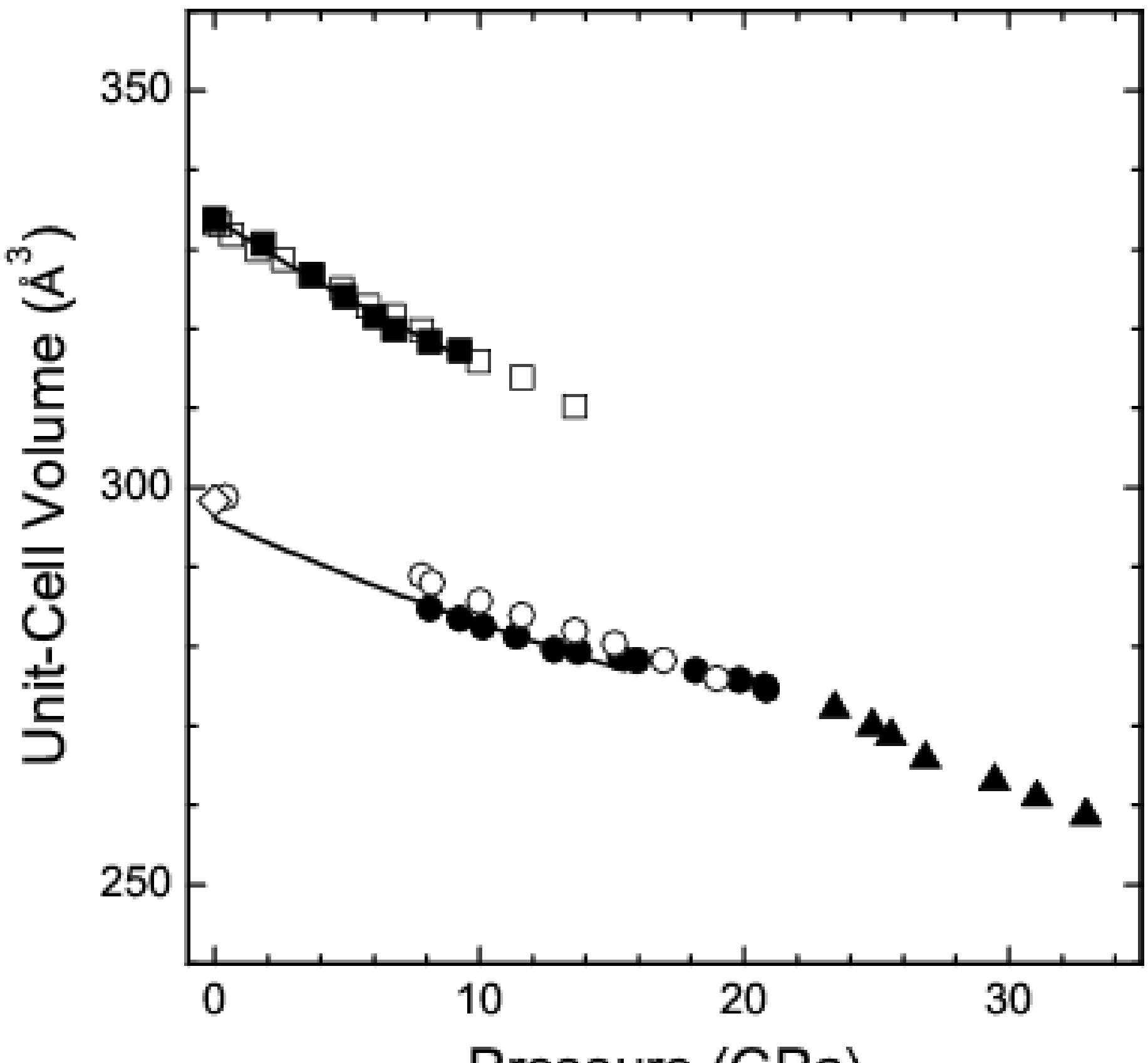